\PassOptionsToPackage{dvipsnames}{xcolor}
\documentclass[10pt,sigconf,nonacm,screen]{acmart}

\usepackage[all]{nowidow}
\usepackage{xcolor}
\usepackage{subcaption}
\usepackage{hyperref}
\usepackage{acronym}
\usepackage[most]{tcolorbox} 
\usepackage{multirow}
\usepackage{fnpct} 

\definecolor{lightgray}{gray}{0.9} 

\usepackage{csquotes}

\usepackage[capitalise,noabbrev]{cleveref}
\usepackage{graphicx}
\usepackage{subcaption}

\crefformat{section}{\S#2#1#3} 
\crefformat{subsection}{\S#2#1#3}
\crefformat{subsubsection}{\S#2#1#3}
\crefrangeformat{section}{\S#3#1#4 to~\S#5#2#6}
\crefrangeformat{subsection}{\S#3#1#4 to~\S#5#2#6}
\crefrangeformat{subsubsection}{\S#3#1#4 to~\S#5#2#6}

\usepackage{todonotes}
\begin{document}





\author{Tim C. Rese}
\affiliation{%
    \institution{TU Berlin}
    \city{Berlin}
    \country{Germany}}
\email{tr@3s.tu-berlin.de}
\orcid{0009-0008-0185-8339}



\author{Nils Japke}
\affiliation{%
    \institution{TU Berlin}
    \city{Berlin}
    \country{Germany}}
\email{nj@3s.tu-berlin.de}
\orcid{0000-0002-2412-4513}

\author{Diana Baumann}
\affiliation{%
    \institution{TU Berlin}
    \city{Berlin}
    \country{Germany}}
\email{diba@3s.tu-berlin.de}
\orcid{0009-0006-3000-6691}

\author{Natalie Carl}
\affiliation{%
    \institution{TU Berlin}
    \city{Berlin}
    \country{Germany}}
\email{nc@3s.tu-berlin.de}
\orcid{0009-0000-5991-9255}

\author{David Bermbach}
\affiliation{%
    \institution{TU Berlin}
    \city{Berlin}
    \country{Germany}}
\email{db@3s.tu-berlin.de}
\orcid{0000-0002-7524-3256}

\title{GeoBenchr: An Application-Centric Benchmarking Suite for Spatiotemporal Database Platforms}


\copyrightyear{2026}
\acmYear{2026}

\begin{abstract}
    The rapid growth of spatiotemporal data volumes needs to be handled by database systems capable of efficiently managing and querying such data. 
    Existing systems such as PostGIS, SpaceTime, and MobilityDB offer partial solutions but differ widely in scope and performance. 
    Also, first spatiotemporal benchmarks provide valuable insights but are limited in scope and, to our knowledge, no application-centric benchmarking suite exists. 

    In this paper, we propose GeoBenchr, an open-source, application-centric benchmarking suite for spatiotemporal platforms.
    GeoBenchr enables comprehensive evaluation across diverse datasets, query types, and workload patterns, reflecting realistic use cases from domains such as aviation and maritime tracking. 
    We use our GeoBenchr prototype to evaluate several system aspects including scalability, configuration impact, and cross-platform performance comparison.
    Our results highlight the importance of application-centric benchmarking in selecting suitable spatiotemporal database systems for real-world scenarios.
\end{abstract}

\maketitle

\section{Introduction}
\label{sec:introduction}
Since the 2010s, the generation of spatiotemporal data has grown rapidly.
Spatiotemporal data contain both spatial and temporal components (commonly represented by geographic coordinates and timestamps), e.g., GPS traces or weather data.
Efficient analysis and real-time processing of such data require database systems  that are capable of handling spatial and temporal dimensions simultaneously.

Specialized database systems, engines, and extensions such as PostGIS\footnote{\url{https://postgis.net/}} and TimescaleDB\footnote{\url{https://www.tigerdata.com/}} have emerged to address these needs, offering tailored query capabilities and optimizations, and have been used to conduct various analyses on spatiotemporal datasets, e.g.,~\cite{gomez2024querying,sakr2023user,natsvlishvili2022development,karakaya2023crowdsensing}.
Combining these functionalities, systems for handling spatiotemporal data have been developed, each with varying features, data formats, and performance characteristics~\cite{hughes2015geomesa,alam2022survey,zimanyi2020mobilitydb}. 
Taking this further, Moving Object Data (MOD) creates unique challenges even for spatiotemporal database systems, as it involves tracking objects that change location over time, leading to complex data structures and query requirements~\cite{guting2005moving}.
Furthermore, as no standard interface, query dialect, or data model exists for spatiotemporal data, the degree of vendor lock-in when choosing a backend is high, making the initial choice of a platform critical for long-term success.
Hence, application developers urgently need comprehensive benchmarks for evaluating and comparing these systems under realistic, application-driven workloads~\cite{bermbach2017book}. 

Existing benchmarking efforts, such as BerlinMOD~\cite{duntgen2009berlinmod}, provide valuable initial insights into spatiotemporal database performance using synthetic datasets and workloads based on vehicle movement data, however, they lack diversity in dataset characteristics and application scenarios.
Previous research, however, has shown that precisely these attributes are key influence factors for performance of spatiotemporal database systems~\cite{rese2025spatbench}. 
Additionally, use cases vary heavily in their required scale and the impact of database configuration is often overlooked in benchmarking studies, despite its strong influence on performance outcomes~\cite{cressie2011statistics}. 

In this paper, we present GeoBenchr, an application-centric benchmarking suite for spatiotemporal platforms, designed to evaluate their performance in realistic, diverse scenarios.
GeoBenchr covers all stages of benchmarking, from data generation to performance analysis, and our prototype currently supports the evaluation of five unique platforms: PostGIS, TimescaleDB, MobilityDB,\footnote{\url{https://mobilitydb.com/}} SedonaDB,\footnote{\url{https://sedona.apache.org/sedonadb}} and the proprietary SpaceTime\footnote{\url{https://www.mireo.com/spacetime}}. 

We make the following contributions:
\begin{itemize}
    \item We design and implement two distinct application-centric benchmark scenarios as part of a comprehensive benchmark suite based on real-world use cases and datasets (\cref{sec:design}).
    \item We provide query translation and client interaction models to allow suite users to easily add their own benchmarks based on novel datasets (\cref{sec:design_trans}).
    \item We evaluate our approach by providing our proof of concept implementation (\cref{sec:impl}) and use the GeoBenchr prototype to comprehensively study performance behavior of spatiotemporal platforms across a range of diverse application scenarios, configurations, and deployments (\cref{sec:experiment-setup}).
\end{itemize}

\section{Background \& Related Work}
\label{sec:background}
This section provides an overview of spatiotemporal data, database systems, and benchmarking approaches.

\subsection{Spatiotemporal and Moving Object Data}
Spatiotemporal data associate an object's location with a timestamp and are generated by sources such as GPS devices and sensors.  

A key distinction lies in whether data stem from stationary or moving devices, leading to moving object data (MOD). 
MOD can be stored either as discrete location updates with timestamps or as reconstructed trajectories~\cite{pelekis2004literature,hamdi2022spatiotemporal}. 
Additional approaches include trajectory segmentation or raster/datacube representations depending on the use case~\cite{singla2021experimental,gao2022multi}.

\subsection{Spatiotemporal Database Systems and Processing Engines}
Spatiotemporal database systems integrate spatial and temporal capabilities, supporting applications such as sensor networks and urban mobility analysis~\cite{bader2017survey,rudakov2023comparison,shekhar2012benchmarking,ray2011jackpine,hulbert2016experimental}.  
Recent distributed systems enable large-scale spatial processing across clusters~\cite{eldawy2015spatialhadoop,yu2015geospark}, while new data formats improve storage and querying efficiency~\cite{wachs2024analysis,butler2014geojson,hu2015describing}.

\begin{figure}[ht]
    \centering
    \includegraphics[width=0.49\textwidth]{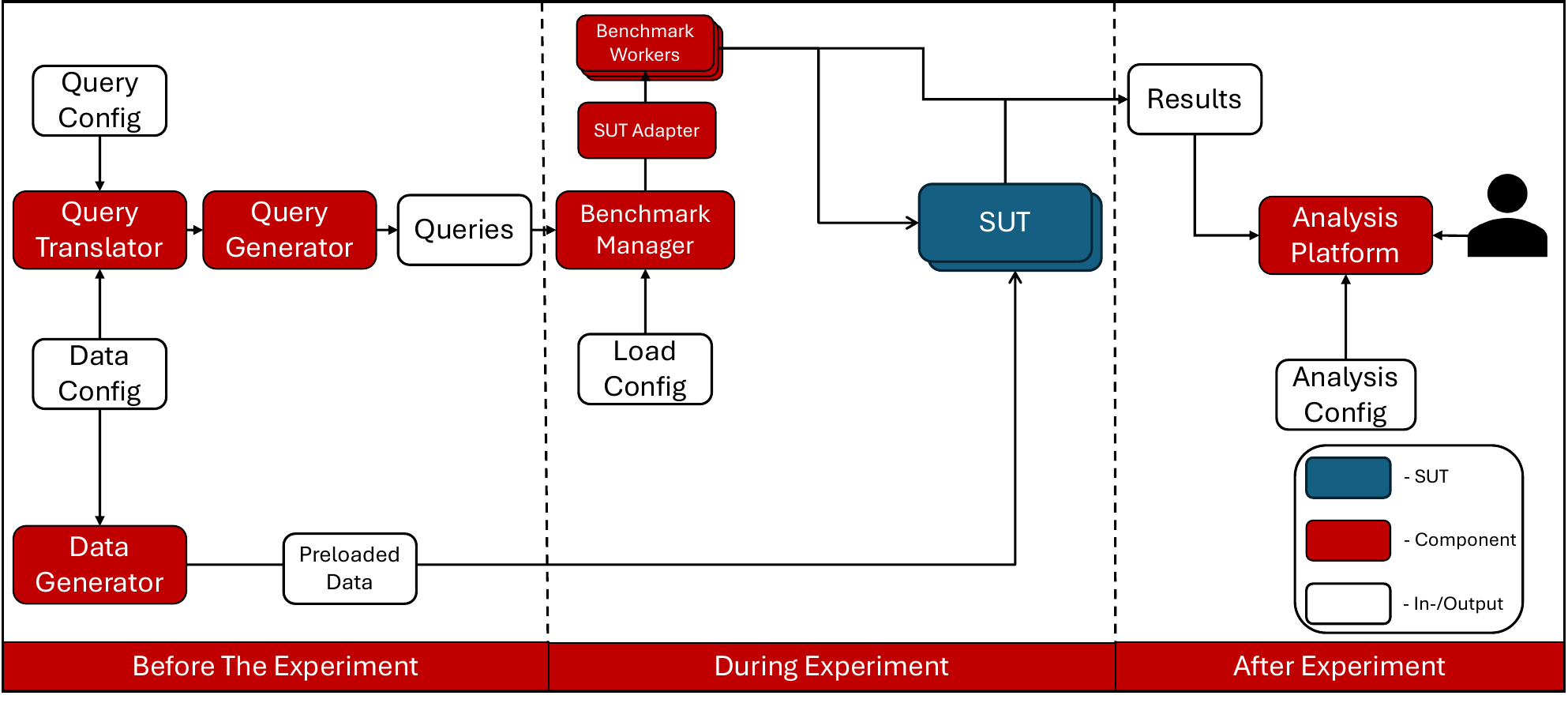}
    \caption{GeoBenchr's modular architecture allows configurable benchmark parameters. Extended from~\cite{rese2025towards}.}
    \label{fig:architecture}
\end{figure}

\subsection{Spatiotemporal Datasets and Generators}
Synthetic generators exist, but benchmarks benefit from real-world datasets and workloads to reflect practical use cases~\cite{duntgen2009berlinmod,kim2023geoycsb}.  
Dataset characteristics such as density and overlap significantly influence performance~\cite{rese2025spatbench}, motivating diverse datasets. 
BerlinMOD defines common MOD query types, including spatial joins, range queries, nearest neighbors, and temporal aggregations~\cite{duntgen2009berlinmod}.

\subsection{Benchmarking Suites}
Benchmarking suites compare Systems Under Test (SUT) using standardized datasets, queries, and metrics.  
General-purpose benchmarks lack spatiotemporal focus and are difficult to adapt, while domain-specific ones (e.g., BerlinMOD, GeoYCSB, SpatialBench) often provide limited dataset variation or application-driven design~\cite{duntgen2009berlinmod,kim2023geoycsb,apacheSpatialBenchSpatialBench}.  
Some approaches focus on index evaluation rather than full system performance~\cite{jensen2006cost,rese2025spatbench}, and prior work typically addresses either cross-system comparison or configuration tuning, but not both~\cite{schoemans2024multi,eltabakh2006space}.  

Comprehensive spatiotemporal benchmarking remains an open challenge. Prior work~\cite{rese2025towards} outlines initial ideas, which we extend in this paper.

\section{System Design}
\label{sec:design}
Designing an application-centric benchmarking suite that can evaluate spatiotemporal platforms across diverse configuration parameters presents several challenges. 
Database systems often store such data in unique formats, requiring different access methods for retrieving the same information. 
Additionally, while a standardized query language such as SQL exists, systems are not bound to use it, leading to different query dialects and functions across platforms. 
As previously mentioned, data heterogeneity within MOD also poses a requirement, as systems may perform differently dependent on dataset characteristics~\cite{rese2025spatbench}.

Here, we provide three separate contributions that address these challenges: First, we design two unique application-centric benchmark scenarios based on real-world use cases and datasets.
Secondly, we propose GeoBenchr as an extensible benchmarking framework in which users can easily add support for additional SUTs as well as additional benchmark scenarios building on other datasets.
Third, to make this feasible, we provide query translation and client interaction models to assert that users of GeoBenchr can focus on application domain aspects such as dataset and queries rather than dealing with query support heterogeneity.
In this section, we describe the design decisions that support these capabilities and discuss how the suite's architecture enables reproducible, fair, and flexible benchmarking of spatiotemporal database systems.
\begin{figure}
  \centering
  \begin{lstlisting}[
      language=SQL,
      basicstyle=\ttfamily\footnotesize,
      frame=single,
      caption={Query Template for AIS.Q1.},
      label={lst:querytemplate}
  ]
  - name: countActiveCrossingsInPeriod
  use: true
  type: temporal
  mobilitydb: |
    SELECT COUNT(*), :period_medium
    FROM crossings c
    WHERE c.traj && :period_medium;
  postgis: |
    SELECT COUNT(DISTINCT c.crossing_id)
    FROM crossing_points c
    WHERE c.timestamp BETWEEN :period_medium;
  sedona: |
    SELECT COUNT(DISTINCT c.crossing_id)
    FROM crossing_points c
    WHERE c.timestamp BETWEEN :period_medium;
  spacetime: |
    SELECT COUNT(*) FROM (
      SELECT DISTINCT crossing_id
      FROM crossing_points c
      WHERE c.t <@ EPOCHRANGE :period_medium
    );
  repetition: 50
  parameters:
  - period_medium
  \end{lstlisting}
\end{figure}  

\begin{figure}
  \centering
  \includegraphics[width=0.4\textwidth]{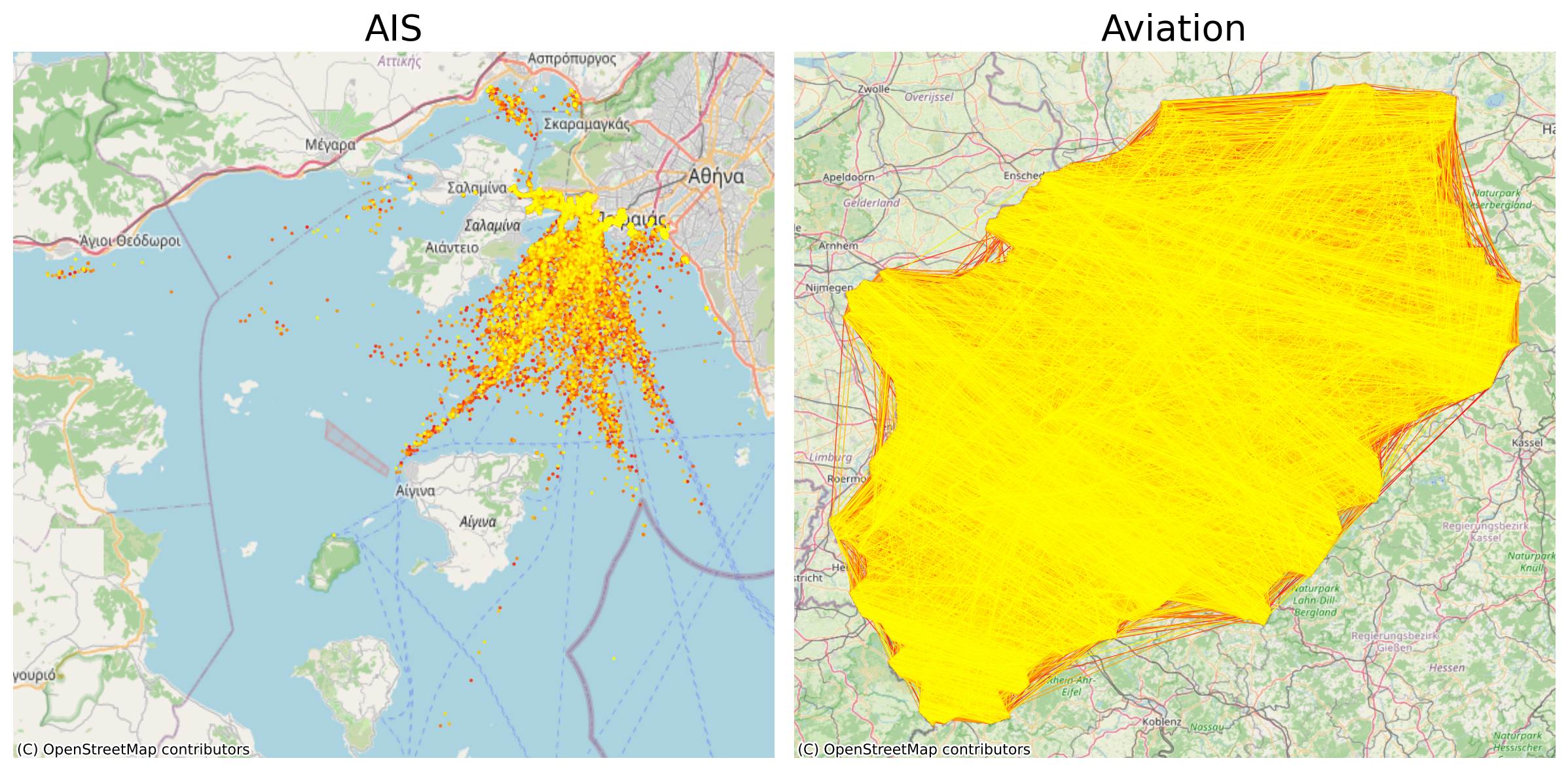}
  \caption{The two real-world MOD datasets we base our application scenarios on, varying heavily in their characteristics such as data distribution and movement patterns. From left to right: AIS data from the published Piraeus AIS dataset~\cite{tritsarolis2022piraeus} and flight data from the Deutsche Flugsicherung (DFS)\protect\footnotemark.}
  \label{fig:datasets}
\end{figure}
\subsection{GeoBenchr Architecture}
GeoBenchr is designed around three key objectives~\cite{rese2025towards}, while considering standard benchmark requirements~\cite{bermbach2017book}:
\begin{enumerate}
    \item \textbf{Workload Flexibility:} Support diverse datasets, query types, and workload patterns to reflect real-world application scenarios.
    \item \textbf{Ease of Extensibility:} Provide a modular architecture that makes it simple to add support for new SUTs.  
    \item \textbf{Custom Analysis:} Include a framework that collects all relevant metrics and allows users to specify which metrics they deem relevant. 
\end{enumerate}

Each of these goals influences specific components of GeoBenchr, including dataset generation, query translation, workload execution, and system configuration.  
GeoBenchr is implemented in a modular fashion, allowing each of these components to be reused or extended independently.
We provide a high-level overview of the GeoBenchr architecture in \cref{fig:architecture}.

\subsubsection*{Dataset Choice and Scaling}
To support (data size) scalability experiments, GeoBenchr includes multiple real-world datasets with distinct properties, all within the domain of MOD, while providing the option to easily extend GeoBenchr with custom datasets, queries, and database systems as well. 
Each dataset can be instantiated in different sizes, which allows suite users to evaluate different scales of database storage size, and we include data generators for each dataset to allow for custom scaling.
We include two datasets with the initial prototype of GeoBenchr, each varying heavily in their use case, object behavior, as well as trip density and length.

Aside from scaling the dataset size, GeoBenchr also allows users to study scale-out behavior of SUTs under different request loads.
This is possible due to its support for distributed deployments enabled by its flexible and scalable worker architecture, enabling multiple workers per instance.
In the rest of this paper, scale and scalability will always refer to scaling data sizes, not request rates or other scalability dimensions.

\subsubsection*{Included Metrics}
Specifically, we measure the end-to-end latency from the client's perspective in our suite.
Additionally, the query throughput, which is the number of queries that can be processed per second, is also tracked. 
In the closed workload model our current prototype uses, both are also interdependent.  
While other metrics such as resource utilization, index build time, and actual storage size are also important, they are not necessarily relevant to the user because they do not affect user experience directly.
Nevertheless, following standard benchmarking best practices~\cite{bermbach2017book}, GeoBenchr also tracks memory and CPU utilization.


\subsubsection*{Query And Data Generation, Translation, and Execution}
\label{sec:design_trans}
We also need to be able to translate queries between different database systems due to the difference in query dialects. 

GeoBenchr includes queries specified in platform-specific dialects, which are parametrized to support varying input parameters. 
To this end, we define spatial, temporal, and spatiotemporal queries in a YAML file, with one example shown in \cref{lst:querytemplate}.

GeoBenchr includes a parameter generator that produces query parameters such as time periods or spatial regions which are based on the properties of the dataset.  
Additionally, we provide options to enable the user to define how many unique parameter sets are generated per query type and how many times each query should be repeated.
\subsubsection*{Benchmark Execution and Parallelism}
Evaluating query performance under both isolated and concurrent workloads is essential for capturing real-world application behavior.  
GeoBenchr supports both \emph{sequential} and \emph{parallel} execution modes, where queries are either run one after another or distributed across multiple concurrent clients.
We follow a closed workload model, where the configuration file indicates how many concurrent users exist during the experiment.
\begin{table*}[t]
  \caption{Selected query set from AIS and Flight applications. A full list of queries for all applications can be found in our online appendix.}
  \centering
  \begin{tabular}{| p{0.12\linewidth} | p{0.12\linewidth} | p{0.12\linewidth} | p{0.12\linewidth} | p{0.12\linewidth} | p{0.12\linewidth} |} \hline
  
  \multicolumn{6}{|c|}{\textsc{Query}} \\ \hline
  
  AIS.Q2: Counts active crossings at a specific hour of day across a given period. &
  
  AIS.Q3: Finds all crossings that passed near a specified island. &
  
  AIS.Q6: Counts unique crossings that passed near a harbor during a given time period. &
  
  Avi.Q1: Counts the number of flight position updates received during a given time period. &
  
  Avi.Q3: Finds all flights that passed through a specified county. &
  
  Avi.Q5: Finds all flights that passed through a specified county during a given time period. \\ \hline
  
  \multicolumn{6}{|c|}{\textsc{Input}} \\ \hline
  
  Hour of day + time period. &
  
  Island name. &
  
  Harbor name + time period. &
  
  Time period (start, end). &
  
  County name. &
  
  County name + time period. \\ \hline
  
  \multicolumn{6}{|c|}{\textsc{Output}} \\ \hline
  
  Number of crossings active at the given hour. &
  
  Set of crossings near the island. &
  
  Number of unique crossings near the harbor in the period. &
  
  Number of flight updates. &
  
  Set of flights intersecting the county. &
  
  Set of flights intersecting the county within the time period. \\ \hline
  
  \end{tabular}
  \label{tab:selected_queries}
  \end{table*}
\subsection{Application-Centric Benchmark Scenarios}
We design two application-centric benchmark scenarios that are based on real-world use cases and user requests. 
Each application scenario includes spatial, temporal, and spatiotemporal queries that reflect common operations within the respective domain. 
To ensure realistic query workloads, we provide supporting datasets that can be used for joins and filtering within the queries.
As an example, we include airport and city locations for the flight application.

Where applicable, we store a trip both as its individual instant as well as a simpler version that includes its start and end location and time, which enables more efficient querying for specific queries, similar to the TLC dataset~\cite{nycTripRecord}.
We provide a visualization of the datasets in~\cref{fig:datasets}.

As part of GeoBenchr, we designed two benchmarking scenarios based on two real-world datasets provided in various sizes as described in Table~\ref{tab:datasets}.
For each of these datasets, we take inspiration from real-world use cases and define six representative queries that reflect possible application requests in each domain.
We then translate these queries into the respective query languages of each system, making sure that the result is the same across all systems for each query.  
We include two MOD datasets in our evaluation, which heavily vary in their characteristics to cover a broad range of possible application use cases.
Each query is either run against the full point dataset or a simplified version that only includes start and end points/times of each trip, where applicable.
\textbf{Q1-2} are temporal queries for each application, \textbf{Q3-4} are spatial queries, and \textbf{Q5-6} are spatiotemporal queries.
We provide literal versions of each query in our online appendix \addtocounter{footnote}{-2}\footnote{\url{https://github.com/timchristianrese/GeoBenchr/blob/main/docs/query-templates.md}}, and exact SQL implementations in our repository.
\subsubsection{\textbf{Flight}}
We consider an application that tracks flights across a certain region, in our case the German state of North Rhine-Westphalia (NRW).
For this, we build on a semi-open source dataset by the Deutsche Flugsicherung (DFS) that includes all flights that passed through the airspace of NRW during 2023. 
These flights include both commercial and private aviation, with varying characteristics such as speed, altitude, and flight path.
For this application scenario, we include further data on districts, counties, cities, and airports in the vicinity of NRW to enable more complex queries.
\cref{tab:selected_queries} provides an overview of the queries that we show in our evaluation, with the full query set being available in our online appendix.

\protect\addtocounter{footnote}{1}
\footnotetext{\url{https://www.dfs.de} - This dataset is closed-source and only available upon request to the DFS.}
\subsubsection{\textbf{AIS}}
We consider an application that tracks vessel movements using AIS data.  
The underlying dataset is kindly provided by Tritsarolis et al.~\cite{tritsarolis2022piraeus} and includes a large amount of vessel trajectories.
We only include a subset of the data in our benchmark scenario and additionally provide island and harbor locations as additional datasets to enable more complex queries. 
\begin{table}
  \centering

  \caption{The dataset sizes included in our benchmark scenarios. We only include entire trips, which leads to the actual number of data points slightly deviating from the exact million.}
  \label{tab:datasets}
  \begin{tabular}{lrrrr}
      \toprule
      Dataset & \# of Points(Mil.) & Trips & Avg. Points / Trip \\
      \midrule
      Aviation & 10 & 47 985 & $\sim$210  \\ 
      Aviation & 100 & 479 846 & $\sim${208}  \\
      Aviation & 257 & $\sim$1.24 Mil & $\sim${207}  \\
      \midrule
      AIS & 10 & 3 057 & $\sim$3 252 \\
      AIS & 92 & 12 814& $\sim$7 221  \\
      \bottomrule
  \end{tabular}
\end{table}

\section{Evaluation}
\label{sec:evaluation}
To showcase GeoBenchr's features, we use it to run various experiments using our proof of concept implementation, which we introduce below.
We show how the benchmark suite can be used to evaluate the scalability of spatiotemporal platforms by benchmarking the supported SUTs with varying data scales.
To conclude our evaluation, we run experiments across all supported SUTs to enable a performance comparison between the different platforms.
In this section, we describe the included query types and applications that we use for all evaluations and the general system setup, and then provide details for each experiment.
\subsection{Proof of Concept Implementation}
\label{sec:impl}
Our proof of concept implementation of GeoBenchr already supports the evaluation of five different SUTs which are currently popular in the spatiotemporal database community, including both in-memory and disk-based systems. 
All design principles described in the previous section are implemented in our prototype. 
\subsubsection{Supported SUTs}
Our goal is to provide a broad overview over currently implemented spatiotemporal/MOD systems, we therefore include a variety of database systems and processing engines in our evaluation.
We show each system in \cref{tab:systems}. 

\begin{table}
\centering
\caption{Overview of evaluated database systems.}
\label{tab:systems}
\begin{tabular}{lll}
    \toprule
    System & Type & Version \\
    \midrule
    PostGIS & Spatial & 3.5.3 \\
    PostGIS \& TimeScaleDB & Spatiotemporal & 3.5.3/2.20.3 \\
    MobilityDB & Spatiotemporal & 1.3.0 \\
    SpaceTime & Spatiotemporal & 2.3.0 \\
    SedonaDB & Spatial & 1.0 \\
    \bottomrule
\end{tabular}
\end{table}
\begin{figure*}[ht]
    \centering
    
    \begin{subfigure}[t]{0.32\textwidth}
        \centering
        \includegraphics[width=\textwidth]{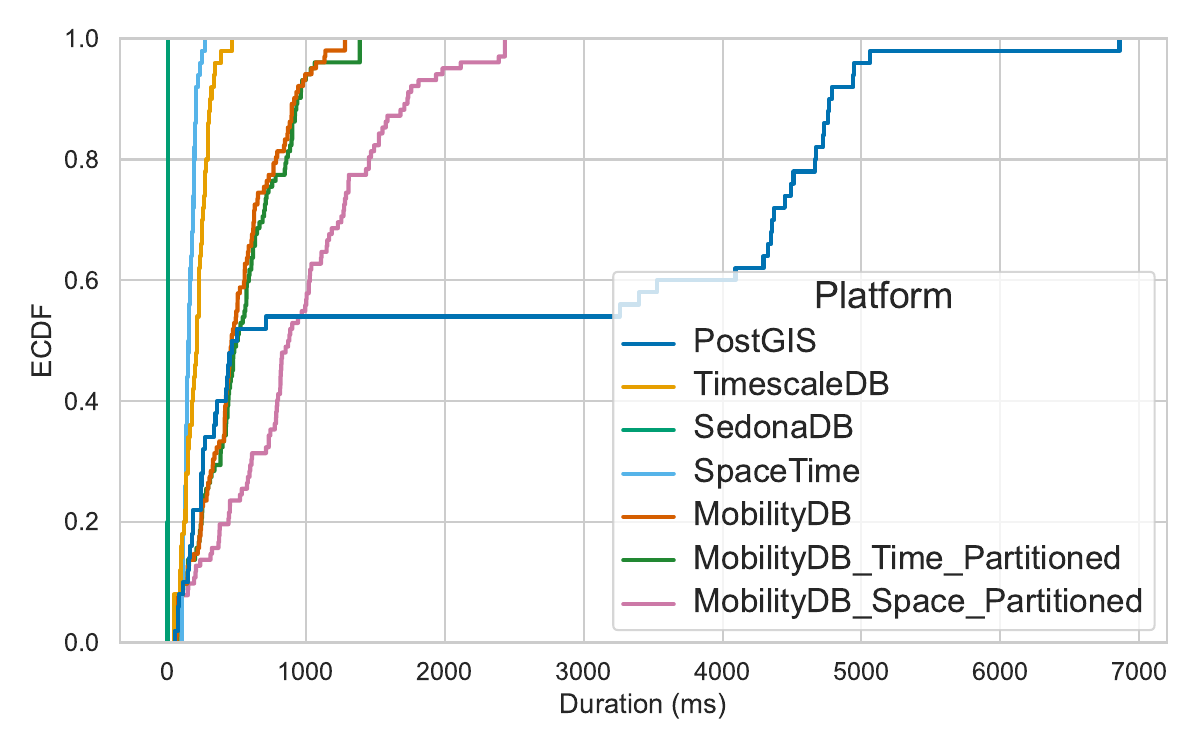}
        \caption{AIS.Q2}
        \label{fig:ecdf-q2}
    \end{subfigure}
    \hfill
    \begin{subfigure}[t]{0.32\textwidth}
        \centering
        \includegraphics[width=\textwidth]{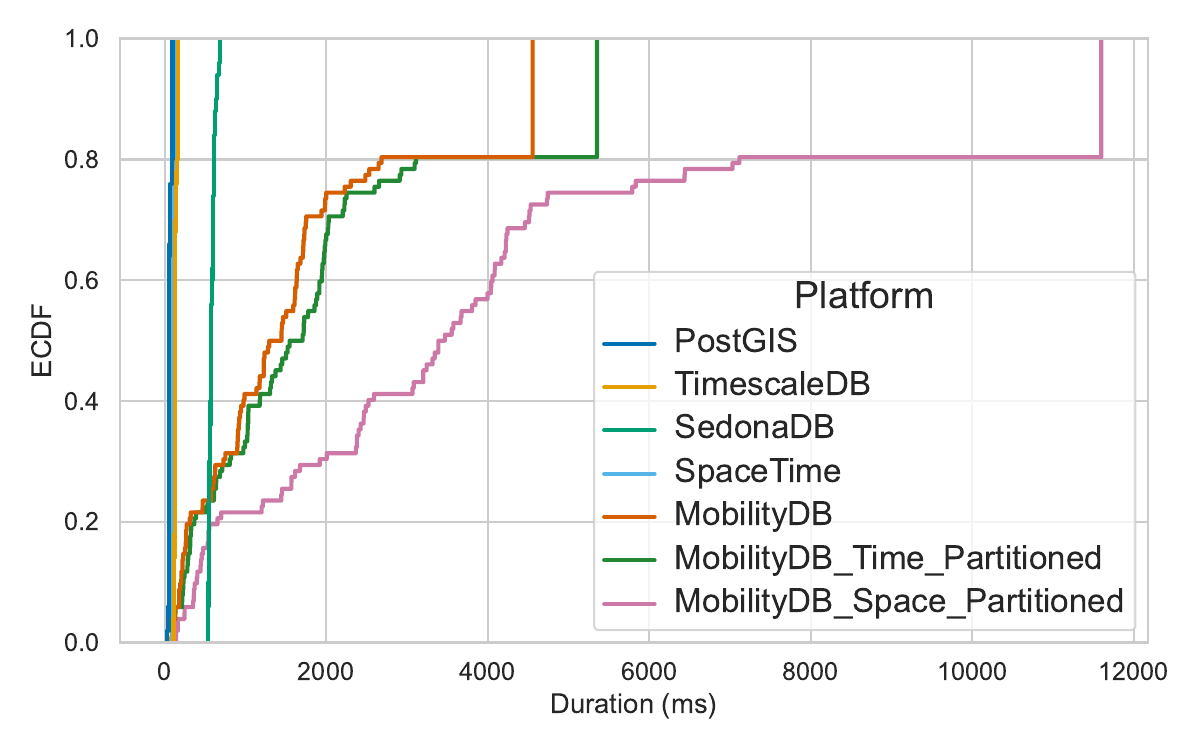}
        \caption{AIS.Q3}
        \label{fig:ecdf-q3}
    \end{subfigure}
    \hfill
    \begin{subfigure}[t]{0.32\textwidth}
        \centering
        \includegraphics[width=\textwidth]{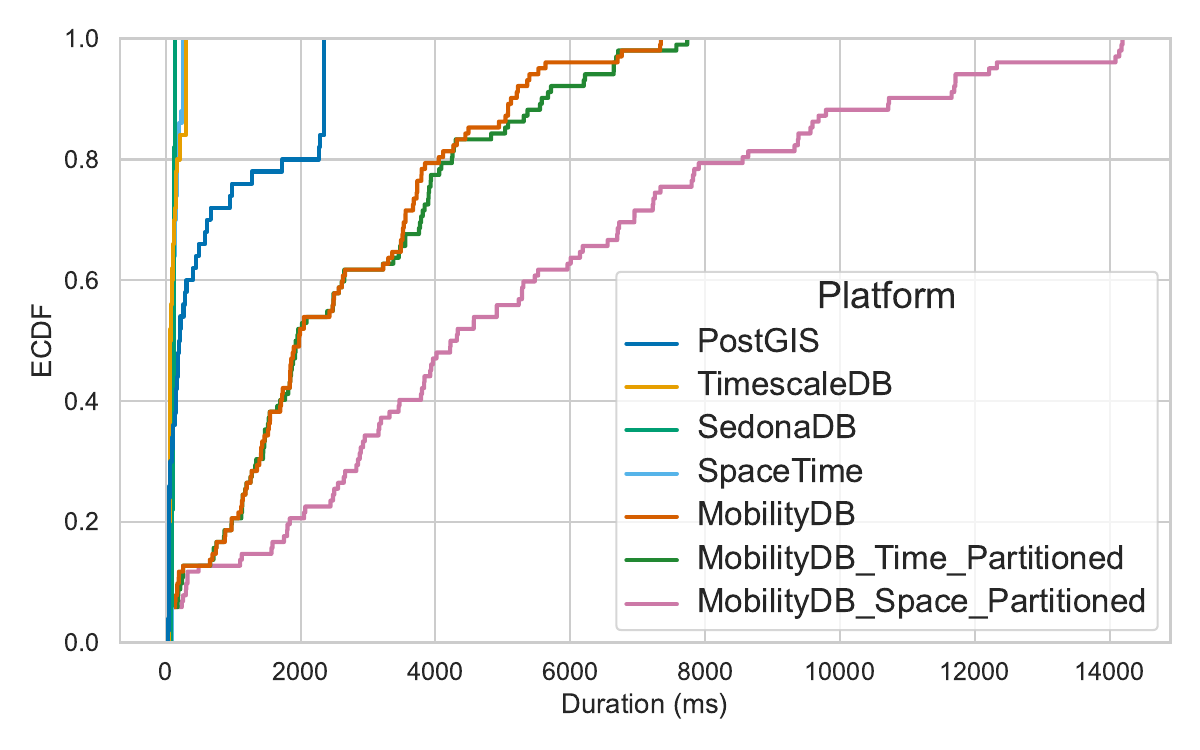}
        \caption{AIS.Q6}
        \label{fig:ecdf-q6}
    \end{subfigure}
    \caption{Empirical Cumulative Distribution Function of Query Durations across datasets. In some cases, database systems manage to outperform SedonaDB despite its in-memory architecture. We also evaluated partitioned MobilityDB versions here.}
    \label{fig:platform-comparison}
\end{figure*}
\begin{figure*}[ht]
    \centering
    \begin{subfigure}[t]{0.32\textwidth}
        \centering
        \includegraphics[width=\textwidth]{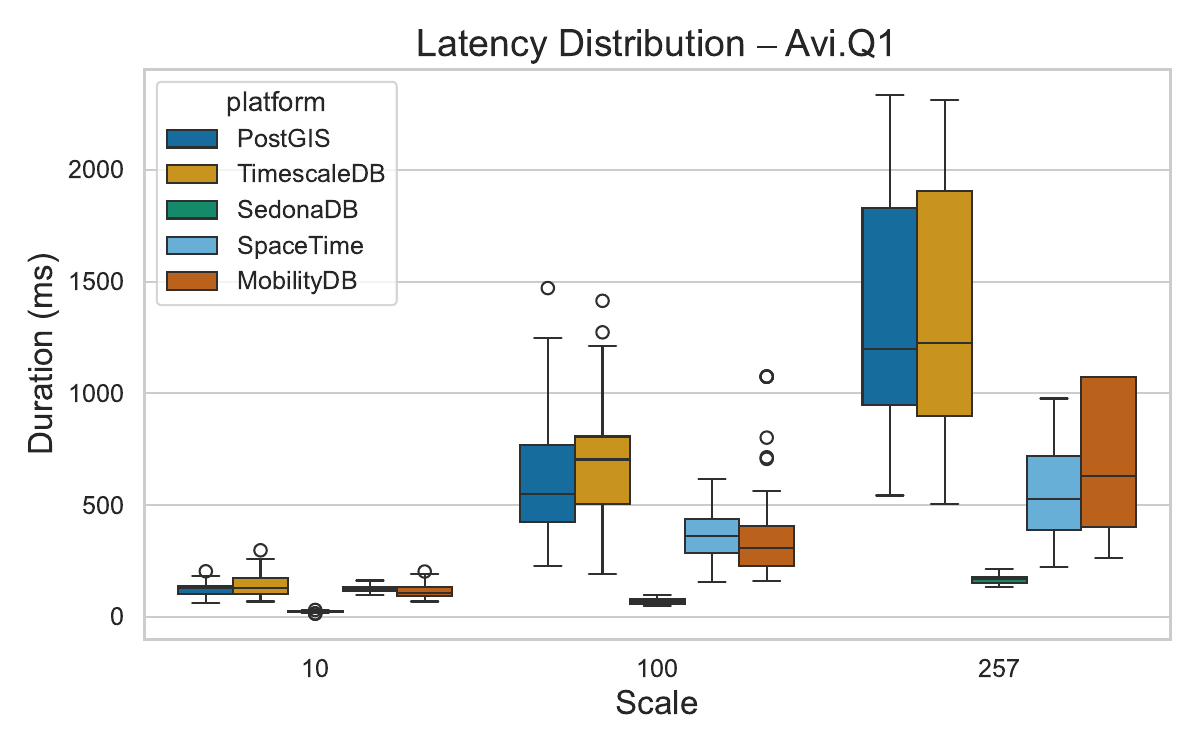}
        \caption{Avi.Q1}
        \label{fig:barplot-q2}
    \end{subfigure}
    \hfill
    \begin{subfigure}[t]{0.32\textwidth}
        \centering
        \includegraphics[width=\textwidth]{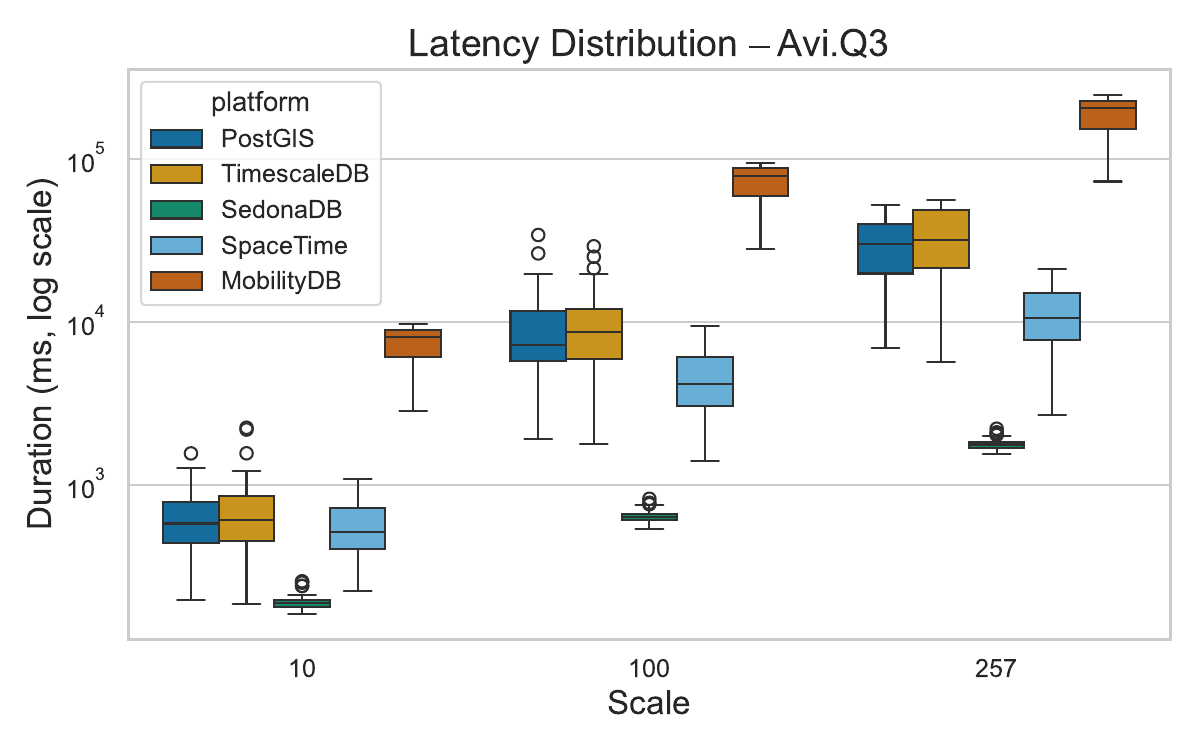}
        \caption{Avi.Q3}
        \label{fig:barplot-q3}
    \end{subfigure}
    \hfill
    \begin{subfigure}[t]{0.32\textwidth}
        \centering
        \includegraphics[width=\textwidth]{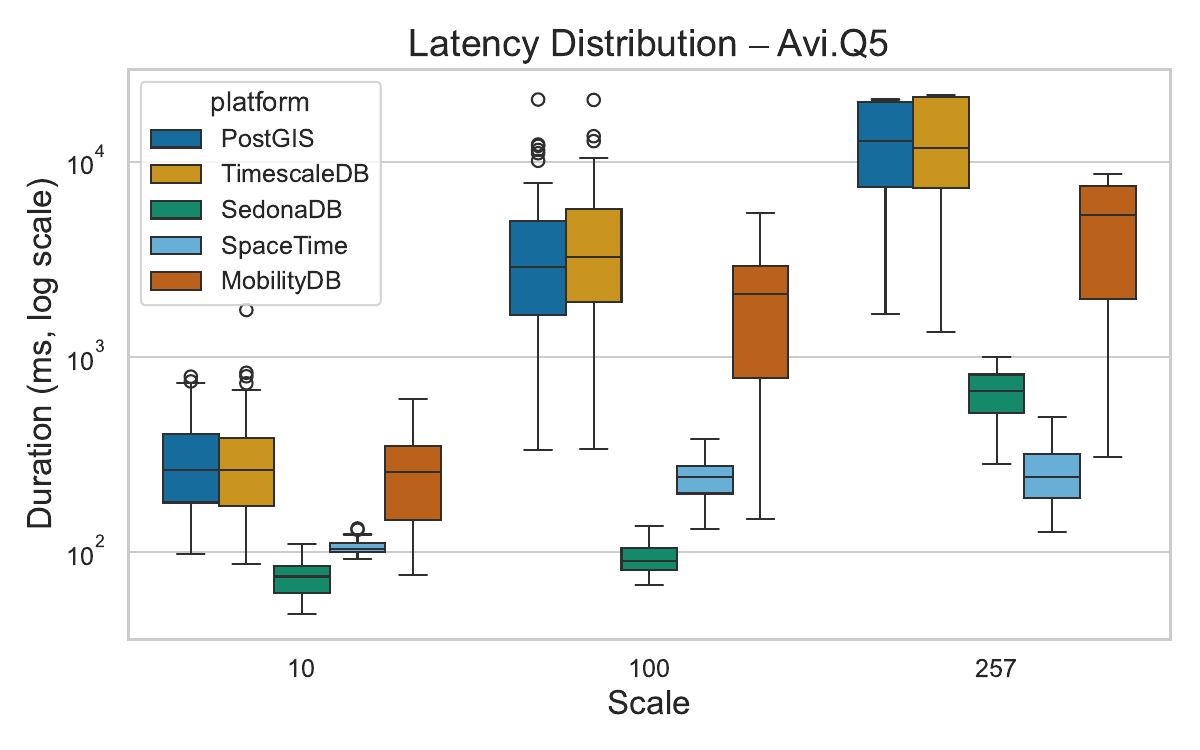}
        \caption{Avi.Q5}
        \label{fig:barplot-q5}
    \end{subfigure}
    \begin{subfigure}[t]{0.4\textwidth}
        \centering
        \includegraphics[width=\textwidth]{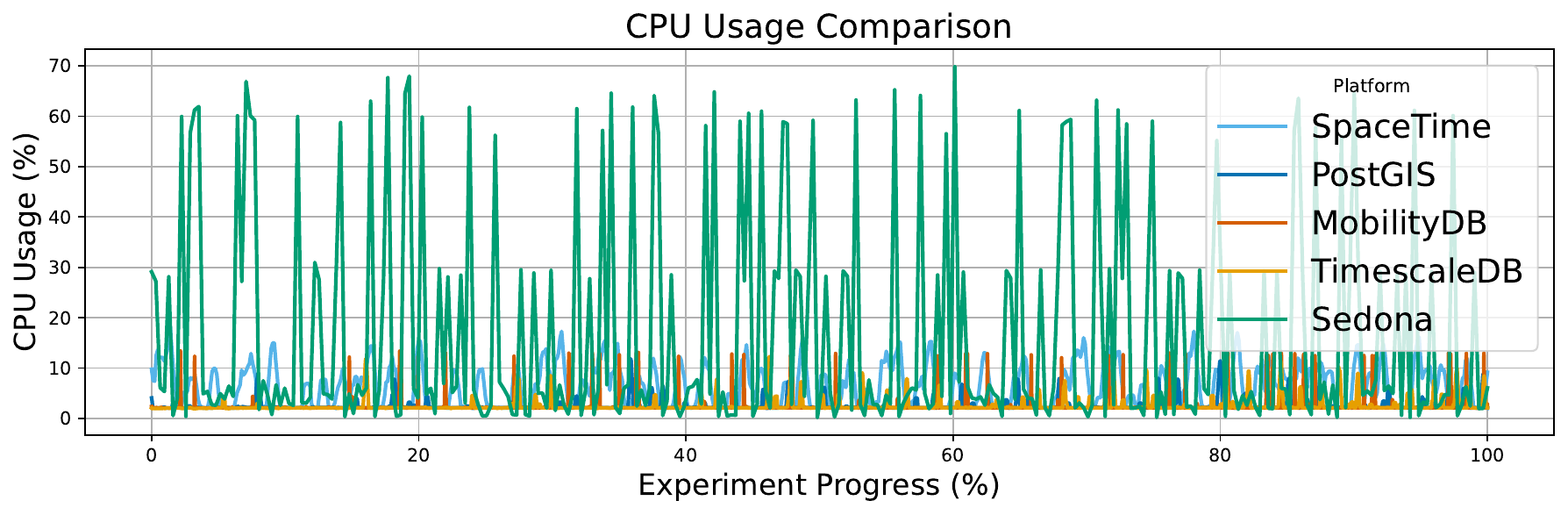}
        \caption{CPU Usage During The Experiment}
        \label{fig:cpu-usage}
    \end{subfigure}
    \begin{subfigure}[t]{0.4\textwidth}
        \centering
        \includegraphics[width=\textwidth]{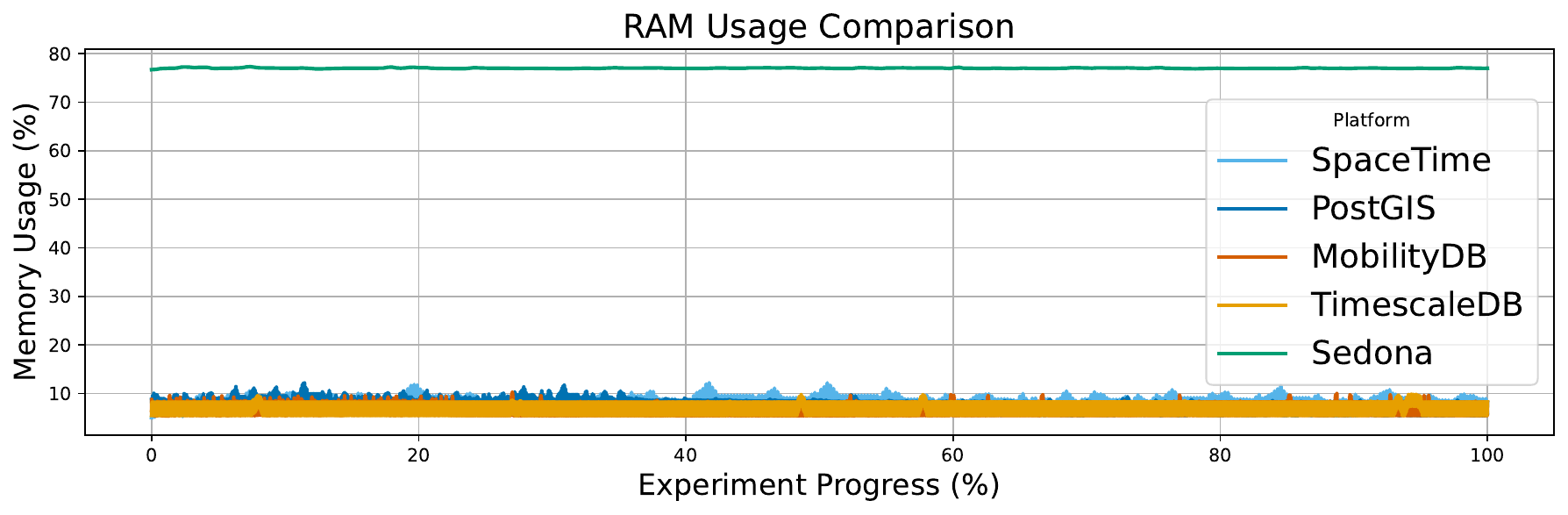}
        \caption{RAM Usage During The Experiment}
        \label{fig:ram-usage}
    \end{subfigure}
    \caption{Depending on the query, different SUTs excel. SedonaDB, while having the best overall performance, is outperformed by SpaceTime for some query/data scale combinations (further shown in \cref{fig:platform-comparison}). MobilityDB shows the advantage it has over PostGIS and TimeScaleDB in some of our queries, while being outperformed in others. Mind the log scale in Avi.Q3 and Avi.Q5. SedonaDB however has a significantly higher CPU and RAM footprint, requiring nearly 77\% of available RAM on average during the experiment, compared to less than 8\% for all other systems.}
    \label{fig:scalabilityresults}
\end{figure*}
\subsection{System and Benchmark Experiment Setup}
\label{sec:experiment-setup}
For each system involved, we follow the best practices with regard to benchmarking and system configuration to the best of our abilities, using documentation where available.
We set up SpaceTime together with its developers, which ensures optimal indexing and configuration for the benchmark. 
We also contacted the MobilityDB developers to ensure that our setup follows their recommendations and we configure other systems according to their respective documentation. 
We host each SUT instance on a 48-core Intel Xeon 4310 CPU with 64GB of RAM, running Debian 13 as all database systems are supported on this platform.
We run each system in isolation on the machine to avoid performance interference, and host the supporting datasets on the same machine to avoid network latency issues.
GeoBenchr itself is hosted on a separate machine to avoid interference with the SUTs.

Each experiment configuration is benchmarked three times in total to account for variability in query execution time.
However, as we will see, our results are quite consistent across runs, with only minor deviations in query latencies. 
Additionally, each system is provided with a warmup phase before each run to reduce outliers caused by cold caches.

Each query, as described above, contains unique parameters to support variability in query execution.
We generate 50 unique parameter sets for each query type, and execute these in random order, i.e., shuffling all queries of all types before execution.
We ensure that the order of execution is the same across all SUTs to ensure comparability of results.

To provide a fair comparison between different systems, we need to ensure that each system is configured optimally for the benchmark workload.
All results can be found in our repository, which also includes our open source benchmarking suite.\footnote{\url{https://github.com/timchristianrese/GeoBenchr}}

\subsection{Scalability Evaluation}
We evaluate the scalability of each system by running GeoBenchr with varying data scales across all SUTs, as described in Table~\ref{tab:datasets}.

\cref{fig:scalabilityresults} shows partial query results for our flight application, with each barplot representing the median query latency for each system at the given data size.
We observe that as long as data can be stored in-memory, SedonaDB provides a significant performance benefit over other systems in most queries. 
However, depending on the query type, SpaceTime actually has lower query latencies than the in-memory solution.
An interesting experiment, which we leave for future work due to paper length restrictions, will be the comparison of distributed SpaceTime and Apache Sedona to see further scalability performance results. 
Since SpaceTime is optimized for large-scale spatiotemporal data, we expect this benefit to become even more pronounced at larger data scales.

\subsection{Cross-Platform Comparison}
We show that GeoBenchr can be used to compare different spatiotemporal platforms by running the same benchmark experiments across all currently supported SUTs. 
For this, we setup each SUT in its largest evaluated data size and run the same benchmark experiments across all systems.
We show some of our results in \cref{fig:platform-comparison}, where only a subset of queries are visualized due to space constraints.
SpaceTime performs well in most of our queries when comparing p99 latencies across SUTs, however, TimeScaleDB and PostGIS also provide competitive performance in several queries.
TimeScaleDB especially can provide a benefit over the default PostGIS setup when considering temporal predicates, as seen in \textbf{AIS.Q3}.
MobilityDB is outperformed by other systems in our evaluation in a high percentage of queries, but provides additional spatiotemporal features that the benchmark here does not visualize, such as additional mobility functions, data storage options and types, and interpolation methods.
The large trip size in some of our datasets make operations on data quite expensive, even when applying indexing and filtering strategies in the query itself. 
SedonaDB shows the best results overall, which is to be expected given its in-memory architecture. 
Even against its closest competitor, its median query latency is 68.38\% faster across all queries in our evaluation.

\section{Discussion \& Future Work}
\label{sec:discussion}
GeoBenchr provides a solid starting point for comparing spatiotemporal database systems, but our prototype and evaluation have limitations that we aim to address in future work.
We outline potential influences on our results and directions for extending GeoBenchr.

\subsection{Discussion}
Several factors may have impacted system performance.
While we aimed for fair treatment, some influences remain.

\paragraph{System Configuration}
We configured each SUT based on best practices and developer input, but cannot guarantee optimal settings across all queries and scales.
Although we included a configuration evaluation, the large tuning space means further optimization could alter cross-system comparisons.

\paragraph{Hardware Influence}
Experiments were conducted on identical hardware for fairness, yet performance may vary across setups.
In particular, systems may behave differently under constrained memory, potentially favoring disk-based approaches.

\paragraph{Dataset and Query Selection}
Our datasets and queries reflect common use cases but do not cover all scenarios.
To address this, we provide translation tools for custom inputs and plan to expand scenario coverage in future work.

\paragraph{SUT Selection}
We compare both in-memory and disk-based systems, though this may disadvantage disk-based systems at smaller scales.
Such systems may perform better on larger datasets that exceed memory capacity.
Traditional benchmarks address this via performance-to-cost metrics, which were not available in our setup.

\subsection{Future Work}
Due to space constraints, we evaluated only five SUTs; GeoBenchr can be extended to include more spatiotemporal, spatial-only, and temporal-only systems.

Our current prototype targets single-node platforms, but extending to distributed systems is straightforward.
Initial experiments indicate minimal overhead, and we plan more comprehensive evaluations.

We also aim to broaden application scenarios, including stationary spatiotemporal data.
This would enable the inclusion of raster database systems, which are well-suited for such workloads.

Finally, while much of the benchmarking process is automated, manual setup and query translation remain.
Ongoing work on a domain-specific language for automatic query translation shows promising results and will be integrated into future versions.
\section{Conclusion}
\label{sec:conclusion}
Spatiotemporal benchmarking suites are essential for evaluating database systems and processing engines, but commonly lack dataset diversity and configurability within the benchmarking process.
To address this, we presented GeoBenchr, an application-centric benchmarking suite of (currently) two application scenarios.
GeoBenchr offers diverse configuration parameters that lets users tune benchmarking experiments to their needs.

We showcased GeoBenchr's capabilities through two large experiments, demonstrating its use for scalability experiments and cross-system comparison, with further results being available online.
Our results highlight the importance of such a suite while also providing insights into the performance of state-of-the-art spatiotemporal data systems.


\balance

\bibliographystyle{ACM-Reference-Format}
\bibliography{bibliography}
\end{document}